\documentclass[%
 reprint,
 amsmath,amssymb,
pra,
]{revtex4-1}

\usepackage{graphicx}
\usepackage{dcolumn}
\usepackage{bm}
\usepackage{xcolor}

\begin{document}


\title{Resistive Hose Modes in Tokamak Runaway Electron Beams}

\author{A.P. Sainterme}
 \email{sainterme@wisc.edu}
\author{C.R. Sovinec}%
\affiliation{%
 Department of Nuclear Engineering \& Engineering Physics, University of Wisconsin - Madison, 1500 Engineering Drive, Madison, Wisconsin  53706-1609 \\
}%

\date{\today}

\begin{abstract}
Beams of energetic runaway electrons are generated during disruptions in tokamaks, and fluid models are used to study their effects on macroscale dynamics. Linear computations of a massless, runaway electron beam coupled to MHD plasma show that resistive hose instabilities grow faster than tearing modes at large resistivity. Eigenvalue results with reduced models of the resistive hose instability are compared with results from the full MHD and beam system, showing that the resistive hose decouples from any plasma response. An estimate of plasma temperature at which growth of the resistive hose dominates tearing for post-disruption DIII-D plasma parameters is in a physically relevant regime
\end{abstract}
\maketitle
The tokamak is the leading candidate for a magnetic confinement fusion reactor. It confines plasma in a toroidal configuration with an externally generated toroidal magnetic field and a poloidal magnetic field supported by toroidal electrical current that is driven through the plasma. Tokamaks are prone to disruptions---sudden dynamic events in which confinement of the plasma is lost and the plasma current is terminated. A typical disruption begins with a rapid loss of plasma thermal energy, a thermal quench (TQ), followed by a current quench (CQ). The decrease in temperature during the TQ increases the electrical resistivity of the plasma, increasing the electric field along magnetic field-lines and accelerating a small population of electrons to high energies. Since the effective collision frequency decreases at higher energies, the resulting beam-like population of supra-thermal electrons is largely unimpeded by collisional drag. These runaway electrons (REs) can carry a substantial fraction of the initial plasma current and have caused considerable damage to plasma-facing components\cite{Breizman2019}. Understanding the dynamics of REs during disruptions is critical for the development of tokamak fusion reactors. For the ITER experiment, which will carry a plasma current of 15 MA, being able mitigate RE-induced damage will be essential\cite{LEHNEN201539}. To that end, the development and characterization of computational tools to simulate RE dynamics in tokamaks is required. 
\par One common method for modeling the macroscale effects of REs in tokamaks treats them as a separate cold beam-like fluid species in extended MHD simulations. The RE beam is treated as a source of resistance-free current density whose direction depends on the time-evolving magnetic field that interacts with a plasma consisting of thermal electrons and ions governed by single-fluid MHD equations. The consistency of this approximation presumes negligible inertia of both electron species, $m_e\ll m_i$, $\gamma_r m_e \ll m_i$, and a small particle density of REs $n_r \ll \rho/m_i $. Here, $\gamma_r$ is the relativistic factor associated with the runaway parallel velocity: $\gamma_r = 1/\sqrt{1-(c_r/c)^2}$.
Since the RE current is essentially resistance free\cite{Breizman2019}, the typical resistive MHD Ohm's law is modified:
\begin{equation*}
    \mathbf{E}+\mathbf{V}\times\mathbf{B} = \eta\mathbf{J} \; \rightarrow \; \mathbf{E}+\mathbf{V}\times\mathbf{B} = \eta\left(\mathbf{J} - \mathbf{J}_r\right).
\end{equation*}
The runaway current density is subtracted from the total current density so that only the fraction of current carried by the bulk plasma is affected by resistivity.
\par Because the resistivity is modified, a natural question is the effect of the inclusion of REs on resistive MHD instabilities. Prior work with this model has explored the effect of a fluid runaway current on the linear behavior of the tearing mode in a sheet pinch \cite{Avinash1988}, and tearing and resistive kink modes in a large aspect ratio cylinder \cite{Liu2020} \cite{Zhao2020}. Helander et al. additionally consider nonlinear saturation of the tearing mode\cite{Helander2007}. 
\par The stability of high-energy, relativistic particle beams propagating through resistive, neutralizing background plasma has also been studied. The earliest analytical description of an instability in such a system is due to Rosenbluth \cite{Rosenbluth1960}. Rosenbluth identifies a kink-type instability of a self-pinched beam of relativistic particles that couples the motion of the beam particles with the resistive response of magnetic field. Weinberg generalizes the work of Rosenbluth first to a spatially modulated beam \cite{Weinberg1964}, where the instability is dubbed a `hose' mode, and later, to a more general dispersion relation for the self-pinched equilibrium \cite{Weinberg1967}  The hose mode has long been recognized in the accelerator community as an impediment to the propagation of beams, and controlling its growth with properly tailored beam profiles has been a subject of study\cite{Fernster1995}.
\par In the resistive hose case, it is assumed that the plasma column surrounding the beam is generated by ionization of a dense gas by the beam particles. The beam is self-pinched, meaning that the self-induced magnetic field is azimuthal, and the equilibrium force balance condition requires that this field is sufficiently strong to provide the centripetal acceleration of the beam particle orbits around the axis. Some relatively recent work includes the effect of an axial magnetic field, which modifies the equilibrium orbits of the beam particles\cite{Uhm1980}. In contrast, in the case of the RE beam generated in a tokamak, the background plasma carries current density before the RE population is generated. There is an existing MHD equilibrium with both axial (toroidal) and azimuthal (poloidal) magnetic field, and after its generation, the RE beam carries some substantial fraction of the equilibrium current density. Despite their differences, we show that important aspects of the resulting mathematical descriptions of the particle-beam and tokamak RE systems are equivalent.
\par In this letter, we draw the connection between the earlier results on resistive hose instabilities of particle beams and resistive instabilities in the context of RE modeling in tokamaks. It is notable that the dynamics of the RE beam in this work are treated under the assumption that the electrons have negligible mass and are guided by the tokamak magnetic fields in contrast to the particle-beam analysis, where inertia is important for the particle orbits. No analysis of the resistive hose in the literature has, heretofore, explored the massless limit. To check its relevance, we estimate that the resistive hose mode that appears in this work would grow faster than the tearing mode at post-TQ plasma temperatures of $T_e\approx 1.5$ eV or less in DIII-D \cite{Lvovskiy2023}.
\par The model consists of a background plasma whose dynamics are governed by single-fluid, resistive MHD. In the present discussion, the effects of pressure are presumed to be negligible, as appropriate for post-TQ conditions. The MHD equations are augmented with a second fluid species that represents the collective low-frequency behavior of REs. The model for REs is an electron beam with a large constant speed parallel to the magnetic field. In the present work, the $\mathbf{E}\times\mathbf{B}$ and curvature drifts are neglected. The RE beam dynamics are coupled to the resistive MHD equations via the modification to the Ohm's law noted earlier. Quasi-neutrality between the species is assumed so that there are no space-charge effects from the beam. 
\par The dynamical equations are linearized about a steady-state solution, and equations are derived for the time-evolution of small perturbations from the equilibrium. \newcommand{\muo}{\mu_0}
\begin{eqnarray}
\label{eq:1}
    \partial_t n_r + \nabla\cdot\left(n_r\mathbf{U}_r + N_r\mathbf{u}_r\right) = 0,\\
    \rho\partial_t \mathbf{v} = \mathbf{j}\times\mathbf{B} + \mathbf{J}\times\mathbf{b},\\
    \partial_t\mathbf{b} = -\nabla\times\mathbf{e}, \\ 
    \nabla\times\mathbf{b} = \mathbf{j}, \\ 
    \mathbf{e} = -\mathbf{v}\times\mathbf{B} + \eta\left(\mathbf{j} - \mathbf{j}_r\right), \\
    \mathbf{U}_r = -c_r\mathbf{B}/B, \\
    \mathbf{u}_r = -c_r\mathbf{b}_\perp/B, \\
    \mathbf{j}_r = -e\left(n_r\mathbf{U}_r + N_r\mathbf{u_r}\right),\\
    \nabla\cdot\mathbf{b} = 0. \label{eq:9}
\end{eqnarray}
Equations \ref{eq:1}-\ref{eq:9} describe the time dependence of the perturbations to the magnetic field, $\mathbf{b}$, the bulk plasma velocity, $\mathbf{v}$, and the runaway density, $n_r$.
The variables with capital letters and the mass density $\rho$ represent the equilibrium fields, and the lowercase variables are the perturbations.
\par As mentioned previously, the impetus for this model is to capture the effect that a presence of a substantial amount of runaway electron current has on instabilities that are present in resistive MHD. An early analytic application of this model to tearing modes in slab geometry found that in the limit of zero inertia, stability of the tearing mode is unaffected\cite{Avinash1988}. More recent results using reduced MHD in cylindrical geometry also suggest that the runaway current does not impact the stability of the tearing mode, but that the nonlinear saturation is affected\cite{Helander2007}. The analytic and numerical work in \cite{Liu2020} finds an additional correction to the linear dispersion relation that results in overstability of the tearing mode.
\par We have implemented equations \ref{eq:1}-\ref{eq:9} in the NIMROD extended MHD code\cite{sovinec2004}. The linear equations are Fourier analyzed in the axial direction, and a high-order spectral element representation is used in the poloidal $(r-\theta)$ plane. For a single axial Fourier harmonic, $e^{ikz}$, the initial value problem is solved by numerically integrating the linear equations in time from an arbitrary initial condition. As time $t \rightarrow \infty$, the solution approaches the most unstable mode from which the growth rate and frequency are determined.
\par For cylindrical problems, the equations can be reduced to a set of coupled ordinary differential equations in the radial coordinate by also assuming a Fourier representation in the azimuthal angle, $e^{im\theta}$ and a complex exponential dependence in time $e^{\gamma t}e^{-i\omega t}$. To verify NIMROD results, a 1D eigenvalue code is used to numerically solve the resulting ordinary differential equations in $r$ to determine $\omega$ and $\gamma$ given $k$ and $m$. The eigenvalue solver also uses spectral elements, albeit for the single spatial dimension $r$\cite{Sovinec2016}.
\par The plasma equilibrium considered here is a $0-\beta$ cylindrical screw pinch with a peaked current density profile. The equilibrium current density is assumed to be entirely carried by the RE beam. This equilibrium approximates the situation observed tokamaks during the current plateau phase that follows a TQ\cite{Hollmann2013}. The safety factor $q$, is given by $q(r) = 1.15(1 + 1.54(r/a)^2)$. Note that $q(0)>1$, and that the profile is monotonically increasing. The only MHD instabilities expected in the absence of runaway effects are tearing modes with $m\geq 2$. 
\par The cylinder is considered periodic in the axial coordinate $z$.  It has unit minor radius, $a=1$, there is a perfectly conducting wall placed at $r=a$, and the RE beam radius is equal to the plasma radius. The axial field strength at $r=0$ is $B_0=1$. The density $\rho$ is chosen so that the Alfv\'en velocity $V_A=1$ when $B_0=1$, and $c_r=20V_A$. This particular value of $c_r$ is chosen as a representative example for $c_r\gg V_A$. The resistivity is normalized to $aV_A$. In the results that follow, the axial wavenumber is $k=-0.1$.
\par
\begin{figure*}[htbp]
    \centering
    \includegraphics{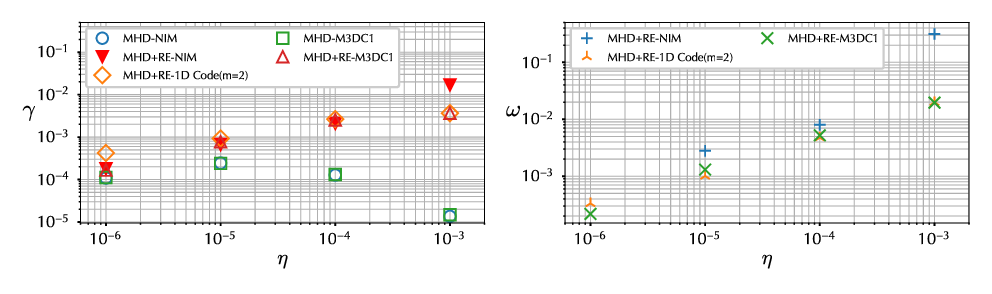}
    \caption{Comparison of the growth rates and frequencies of the fastest growing mode computed by NIMROD, M3D-C1, and the 1D eigenvalue code ($m=2$). In the  NIMROD result, the resistive hose mode is the fastest growing at $\eta=10^{-3}$.}
    \label{fig:nim_m3d_comp}
\end{figure*}

\begin{figure}[htbp]
    \centering
    \includegraphics{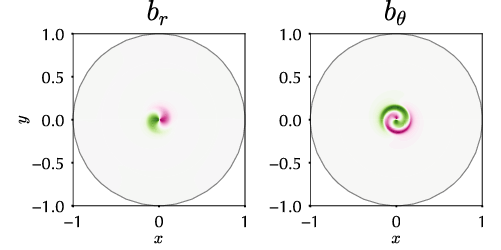}
    \caption{Contours of the radial and azimuthal components of the perturbed magnetic field associated with the growing mode observed at $\eta = 10^{-2}$.}
    \label{fig:remhd_hosemode_con}
\end{figure}

 Figure \ref{fig:nim_m3d_comp} plots the growth rate and frequency of the fastest growing mode from NIMROD calculations of equations \ref{eq:1}-\ref{eq:9} as a function of varying resistivity values. Also plotted are the growth rates of the fastest growing mode in the absence of the RE beam, growth rates and frequencies computed with the M3D-C1 code\cite{Liu2020}, and growth rates and frequencies computed with our 1D eigenvalue code with $m=2$.   
For values of $\eta\lesssim 10^{-4}$, the fastest growing mode is the $m=2$ tearing mode, which displays the expected localization around the $q=2$ surface. For $\eta > 10^{-4}$, the fastest growing mode computed in NIMROD is a hose-like mode with $m=1$. The appearance of this hose mode accounts for the distinct growth rate and frequency at $\eta=10^{-3}$ reported by NIMROD. Figure \ref{fig:remhd_hosemode_con} plots contours of the radial and azimuthal components of the solution for $\eta = 10^{-2}$.
\par
\begin{figure}[htbp]
    \centering
    \includegraphics{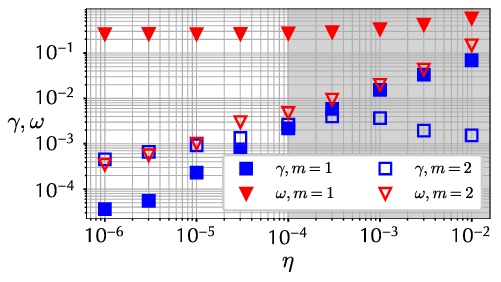}
    \caption{Growth rate and frequency of the fastest growing mode vs. resistivity from the full MHD eigenvalue calculations for the $m=1$ beam mode and the $m=2$ tearing mode.}
    \label{fig:remhd_hosemode_m1_m2}
\end{figure}

 The transition from the tearing mode to the hose mode is explicated by calculating each $m$ separately with the eigenvalue code. Figure \ref{fig:remhd_hosemode_m1_m2} plots the frequencies and growth rates for $m=1$ and $m=2$ as a function of resistivity. The shaded region denotes the range of $\eta$ values where the hose mode grows faster than the tearing mode.
It is clear from the eigenvalue code results that the hose mode scales linearly with resistivity for small $\eta$, whereas the tearing mode scales as $\eta^{3/5}$. The difference in scaling suggests that for equilibria that are unstable to both modes, there is a value of $\eta$ at which the fastest growing mode transitions from tearing to the hose. The exact details of the transition depend on the equilibrium current profile, and quantifying this will require an analytic expression for the hose dispersion relation in this model. Deriving such an expression remains a task for future work.
\par Since the resistive hose mode is driven by the interaction of the beam with the magnetic field, we can isolate the instability from the full system by simply neglecting the MHD velocity perturbation $\mathbf{v}$. This is a good approximation when the background plasma response is on a much longer time scale than the hose mode dynamics. In making this simplification we also redefine the RE density variables as
\begin{equation}
    \lambda \equiv \frac{ec_rn_r}{B},
\end{equation}
and
\begin{equation}
    \Lambda \equiv \frac{ec_rN_r}{B}.
\end{equation}
Then, when $\mathbf{v}=0$, we can consider only the equations
\begin{eqnarray}
   { \partial_t \lambda - c_r\frac{\mathbf{B}}{B}\cdot\nabla\lambda = \frac{c_r}{B}\nabla\cdot\left( \Lambda\mathbf{b}_\perp\right) \label{eq:rebc1}}, \\
    {\partial_t \mathbf{b} - \eta\nabla^2 \mathbf{b} = \eta\nabla\times\left(\lambda\mathbf{B} + \Lambda \mathbf{b}_\perp\right)\label{eq:rebc2}}.
\end{eqnarray}
Eigenvalues for equations \ref{eq:rebc1} and \ref{eq:rebc2} are computed with the same 1D eigenvalue code. Note that in this system of equations, $V_A$ is not a relevant parameter.
\par
\begin{figure}[htbp]
    \centering
    \includegraphics{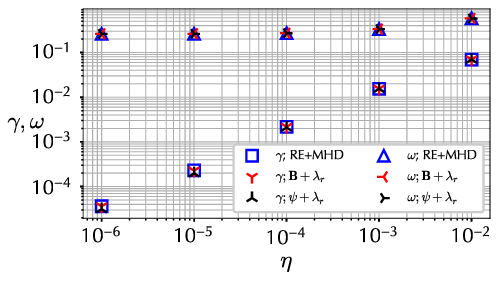}
    \caption{Growth rates and frequencies of the hose mode $m=1$ computed with the full MHD, the model without bulk plasma flow velocity (eqs. \ref{eq:rebc1}-\ref{eq:rebc2}), and with $\mathbf{b}=\nabla\psi\times\hat{\mathbf{z}}$ (eqs. \ref{eq:psi}-\ref{eq:lambda}).}
    \label{fig:psil_rebc_remhd}
\end{figure}

 For the particular equilibrium considered, both with the full MHD system, and with $\mathbf{v}=0$, the polarization of the perturbed magnetic field for the hose-like mode is primarily transverse to the axis of the cylinder. \textit{i.e.}, the axial component of the solution for the perturbed magnetic field is negligible: $\mathbf{b}\cdot\hat{\mathbf{z}} \approx 0$. This observation motivates the introduction of a Clebsch representation for the perturbed magnetic field, $\mathbf{b} = \nabla\psi \times \hat{\mathbf{z}}$ .
 In this approximation, again assuming the dependence $\psi = \psi(r)\exp(im\theta+ikz)$, equations \ref{eq:rebc1} and \ref{eq:rebc2} reduce to two scalar equations for the complex quantities $\psi$ and $\lambda$:
    \begin{eqnarray}
         \partial_t \psi - \eta \nabla^2\psi &=& \eta\left(B_z\lambda + \Lambda\frac{B_\theta B_z}{B^2}\psi'\right)  \nonumber \\
           &  & + \eta\frac{kr}{m}\left(1-\frac{B_\theta^2}{B^2}\right)\psi'\label{eq:psi},
   \end{eqnarray}
   \begin{eqnarray}
    \partial_t\lambda - ic_rF\lambda &=& \frac{c_r}{B}\Lambda'\frac{im}{r}\psi + ic_rF\Lambda\frac{B_\theta}{B^2}\psi',
    \label{eq:lambda}
    \end{eqnarray}
where $F\equiv (mB_\theta/r + kB_z)/B$. Assuming an exponential time dependence results in a coupled eigenvalue problem with a second-order radial derivative operator acting on $\psi$. As shown in figure \ref{fig:psil_rebc_remhd}, numerical solutions of these equations confirm that they are sufficient to reproduce the features of the hose instability observed in the solutions of equations \ref{eq:1}-\ref{eq:9}.
\par The agreement between these three calculations confirms that the flow of the background plasma is unimportant and that the polarization of the magnetic field is predominantly transverse. Moreover, the form of equation \ref{eq:psi} is equivalent to the field equation derived by Rosenbluth and Weinberg to describe the resistive hose mode\cite{Rosenbluth1960}\cite{Weinberg1967}.
\par
\begin{figure}[htbp]
    \centering
    \includegraphics{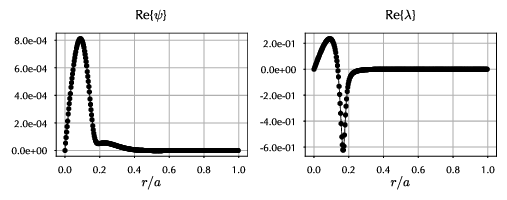}
    \caption{Radial profiles of the most unstable eigenfunction of equations \ref{eq:psi} and \ref{eq:lambda} for $(m=1,k=-0.1)$ and $\eta=10^{-2}$.}
    \label{fig:psil}
\end{figure}

 The hose mode in these calculations does not extend beyond a given radius. Figure \ref{fig:psil} shows radial profiles of the hose mode. It is clear that there is some radius $\tilde{r}$ outside which the profiles of the solution for $\lambda$ and $\psi$ drop to zero. The value of $\tilde{r}$ appears to be determined by the solution to $\omega + c_rF(\tilde{r}) = 0$.
Although there is no rational surface in the tearing sense of $F(r)=0$, the equation for $\lambda$ becomes singular in the rotating frame of the eigenmode at the radial location where $\omega + c_rF(r) = 0$.

We have shown that resistive hose-type modes associated with a RE beam interacting with a thermal background plasma grow faster than the typical resistive MHD tearing mode at large resistivity. The scaling of the hose mode is linear with the resistivity in the low resistivity regime, and the frequency of the mode is much larger than the growth rate. Since the hose mode is dominant when the background plasma resistivity is high, it may appear in post-TQ tokamak plasmas. 
\par For the RE beam profile considered here, the hose mode grows faster than the tearing mode number at a normalized resistivity of $\eta/aV_A \approx 2\times 10^{-4}$. Ignoring for now whether or not this current density profile is representative of the experiment, we may use the plasma parameters from a DIII-D discharge reported in Reference \cite{Lvovskiy2023} to estimate the temperature where this transition occurs. Using $a=0.5, V_A\approx 6.6\times 10^6$, and $Z_{eff}=3$, the Spitzer formula produces a temperature estimate of $T_e \approx 1.5$eV. This is in the regime of temperatures reported for the post-TQ plasmas observed in DIII-D. Even in cases where the hose mode is subdominant to resistive MHD instabilities, it can introduce an $m=1$ component into the fluctuation spectrum that would be otherwise unexpected from resistive MHD alone in cases where the minimum $q$-value is larger than unity.
\par Future work remains to find cases in which an analytic dispersion relation may be found to compare with numerical results. We also intend to address the nonlinear dynamics of the hose mode and more thoroughly characterize its influence on RE beam formation and termination in tokamaks.

\begin{acknowledgments}
We wish to thank Dr. Chang Liu for providing data on the tearing mode results with runaways.
This work is supported by the US DOE through grant DE-SC00180001.
\end{acknowledgments}

\appendix


\bibliography{refs}

\end{document}